\documentclass{aip-cp}

\usepackage[numbers]{natbib}
\usepackage{rotating}
\usepackage{graphicx}
\usepackage{epstopdf}
\usepackage{combelow}
\usepackage{color}
\usepackage{caption}
\usepackage{subcaption}

\def\feq{\ensuremath{f^{(\mathrm{eq})}}}
\def\wp{{\overline{p}}}
\def\vec#1{\mathbf{#1}}
\def\eqref#1{(\ref{#1})}

\begin{document}

\title{Quadrature-based Lattice Boltzmann Model \\for Relativistic Flows}

\author{Robert Blaga}
\eaddress{robert.blaga@e-uvt.ro}
\author{Victor E. Ambru\cb{s}}
\eaddress{victor.ambrus@e-uvt.ro}

\affil{Department of Physics, West University of Timi\cb{s}oara, \\
 Bd. Vasile P\^ arvan 4, Timi\cb{s}oara 300223, Romania}

\maketitle

\begin{abstract}
A quadrature-based finite-difference lattice Boltzmann model is developed that is suitable for simulating relativistic
flows of massless particles. We briefly review the relativistc Boltzmann equation and present our model. The quadrature is constructed such that
the stress-energy tensor is obtained as a
 second order moment of the distribution function. The results obtained with our model
are presented for a particular instance of the Riemann problem (the Sod shock tube).
We show that the model is able to accurately capture the behavior across the whole domain of relaxation times, from the
hydrodynamic to the ballistic regime. The property of the model of being extendable to arbitrarily high orders is shown
to be paramount for the recovery of the analytical result in the ballistic regime.
\end{abstract}

\section{INTRODUCTION}
Relativistic fluid dynamics is an active and dynamical field of current research. The main areas of application are the arena
of astrophysical phenomena \cite{Rezzolla} and high energy nuclear collisions (e.g quark-gluon plasma) \cite{RomatschkeReview}.
A plethora of numerical methods have been developed for solving the relativistic variant of the (macroscopic) hydrodynamical
conservation equations (see, e.g., Ref.\cite{Rezzolla} and references therein).
An alternate approach that has become a popular choice in the past decades for non-relativistic fluid dynamics and 
recently for relativistic flows is to solve instead the Boltzmann kinetic
equation. In this approach the dynamical quantity is a one-particle distribution function, while the macroscopic quantities are
obtained as moments of the distribution function. The non-linearity in the macroscopic equations is traded for the complexity
of the phase space employed in the mesoscopic description, which in turn is mitigated by choosing efficient discretization
schemes. These are called Lattice Boltzmann (LB) models. This approach has the advantage of mathematical simplicity, computational
efficiency and the ability to handle complex geometries \cite{Succi}. Furthermore, by using this description at the level of the
Boltzmann equation, we can accurately describe the evolution of nonequilibrium systems and of rarefied gases, where the hydrodynamic
description is no longer applicable. \par
Traditional lattice Boltzmann methods employ the collision-streaming paradigm \cite{Succi}. The essence of the method
is to choose a finite set of momentum vectors (or equivalently, velocities) such that the constituents of the fluid 
move along the links of the spatial lattice.
The set of velocities must have sufficient symmetry in order to exactly recover the macroscopic moments up to a desired order.
To access physics beyond the hydrodynamic regime, higher-order moments must be recovered. The collision-streaming method is computationally
very efficient, but has the flaw of being difficult to extend in terms of the number of velocities.
In conventional on-lattice collision-streaming models, this can be achieved by either using a larger set of velocities, implying
jumps over an increasing number of neighbours on the spatial lattice \cite{philippi06}, or by using multiple distribution
functions \cite{lallemand03}. If we relax the requirement of having on-lattice velocities, we exchange some of the computational
benefits for the freedom
of choosing large sets of velocities.

The Lattice Boltzmann models have been very popular for describing nonrelativistic fluid dynamics. In recent years, different
LB models have also been developed for describing relativistic flows \cite{Miller10,Miller10.2,Miller11,Miller13, MillerSpheric}.
In this paper, we employ the powerful technique of quadratures to select the set of momentum vectors. This approach has the
advantage of being readily extendable to arbitrarily high orders of the quadrature, allowing the recovery of higher-order moments,
thus granting access to the physics beyond the hydrodynamic regime. Since typically, the resulting velocity sets are
off-lattice, the advection in such models can no longer be performed using an exact streaming step, hence any of the powerful
finite difference, finite element or finite volume methods developed for hyperbolic equations may be employed \cite{Rezzolla}. 

Most of these relativistic LB models (RLB) are of the collision-streaming type.
 We test our model on a particular case of the 1-dimensional Riemann problem, called the Sod shock-tube \cite{Rezzolla}.
 The setup consists of
two
fluid domains
at rest, having different pressures and densities, separated,
 e.g.,
by a thin membrane. When the membrane is removed, a shock-wave and a rarefaction wave propagate in opposite directions.
Due to these features, the Riemann problem is a challenging trial for numerical methods.

\section{RELATIVISTIC LATTICE BOLTZMANN EQUATION}

In order to simulate the relativistic flow of massless particles on Minkowski space-time,
we use the relativistic Boltzmann equation \cite{Cercignani02}:
\begin{equation}
 p^\mu \frac{\partial f}{\partial x^\mu} = \frac{p^\mu u_\mu}{\tau} (f - \feq),
\label{Boltz_gen}
\end{equation}
where the right hand side represents the Anderson-Witting approximation for the collision term
\cite{AndersonWitting}, while $p^\mu$ is the particle four-momentum, obeying the mass-shell
condition\footnote{In this paper we employ the signature $(-, +, +, +)$ for the Minkowski metric,
as well as
Planck units, such that $c = K_B = 1$.}
$p^\mu p_\mu = -(p^0)^2 + (p^x)^2 + (p^y)^2 + (p^z)^2 = 0$.
In this paper, we consider the relaxation time $\tau$ to be constant.
Furthermore, we only consider the simple setup of the one-dimensional Riemann problem in flat spacetime, such that the
distribution function can be
taken to be homogeneous with respect to the perpendicular directions $x$ and $y$.
The Boltzmann equation thus simplifies to:
\begin{equation} \label{Boltz}
\left(\partial_t + v^z \partial_z \right) f = -\frac{u^0 - v^z u^z}{\tau}\left(f - \feq\right),
\end{equation}
where $v^z \equiv p^z / p^0$ is the particle velocity along the $z$ axis.
The equilibrium distribution function is taken to be the Maxwell-J\"uttner distribution for massless particles:
\begin{equation} \label{feqMJ}
\feq = \frac{n}{8\pi T^3}\exp\left(\frac{p^\alpha u_\alpha}{T} \right).
\end{equation}
The
 hydrodynamic fields which describe the macroscopic state of the fluid are obtained as moments of the distribution function:
\begin{eqnarray}
N^\mu  = \int \frac{d^3p}{p^0}\, p^\mu\, f(t,\vec{x},\vec{p}), \quad \qquad 
T^{\mu\nu} = \int \frac{d^3p}{p^0}\, p^\mu p^\nu\, f(t,\vec{x},\vec{p}).
\end{eqnarray}
The equilibrium particle 4-flow and stress-energy tensor (SET) corresponding to \eqref{feqMJ} is of the perfect fluid form:
\begin{equation}
\ \ N^\mu_{\rm eq} = n u^\mu, \qquad \qquad \qquad \qquad \quad \ 
T^{\mu\nu}_{\rm eq} = \left(\epsilon + P\right)u^\mu u^\nu
+
P g^{\mu\nu}, \quad \qquad \quad
\end{equation}
where $n$ is the particle number density, $\epsilon$ is the energy density and $P = nT$ is the hydrostatic pressure, written
in terms of the macroscopic temperature $T$.
The macroscopic velocity of the fluid $u^\mu$ is defined such that it reinforces the conservation of the SET.
It can be seen by multiplying (\ref{Boltz_gen}) with $ p^\nu$ and integrating with the appropriate measure,
that the correct choice is determined by the Landau-Lifshitz condition \cite{MillerSpheric,AndersonWitting,landau87}:
\begin{equation}
T^\mu{}_{\nu} u^\nu = -\epsilon u^\mu.
\end{equation}
\section{LATTICE BOLTZMANN MODEL}

At the core of the lattice Boltzmann method lies the discretization of the momentum space, which is performed such that
certain moments of the equilibrium distribution function $\feq$ are exactly recovered. Achieving this goal is a two-step procedure:
first, a quadrature procedure must be defined by means of which the integrals over the momentum space can be performed exactly;
the second step consists in replacing the collision term by a finite polynomial compatible with the quadrature scheme, such that its
moments are exactly recovered. The details of these two steps can be found in the subsequent subsections.

\subsection{Expansion of the equilibrium distribution function}

In order to perform an expansion of $\feq$, Eq.~(\ref{feqMJ}) can be cast as follows:
\begin{eqnarray} \label{feq}
\feq = \frac{n}{8\pi T^3}\exp\left( \frac{\bar{p}}{\theta}(u_0 - \vec{v}\cdot\vec{u})\right),
\end{eqnarray}
where $\theta = T/T_0$, $\bar{p} = p^0/T_0$ and $\vec{v} = \vec{p}/p$, with $T_0$ being a reference temperature.
The form (\ref{feq}) lends itself to a decomposition
with respect to spherical coordinates, 
similar to that 
performed in Ref.\cite{VictorSpheric} for the nonrelativistic case. In Ref.\cite{MillerSpheric}, such a
decomposition is performed also for the relativistic case, using generalized Laguerre polynomials
$L^{(3)}_{\,l}(\wp)$ of order 3 for the radial component $p$ and vector polynomials
$P^{\,(n)}_{i_1...i_n}(\vec{v})$ for the angular part, which is specifically designed to recover the stress-energy tensor.
We perform here a similar decomposition, but using generalized
Laguerre polynomials of order 1, allowing us to build quadratures that grant access also to the particle 4-flow $N^\mu$,
as opposed to just the SET $T^{\mu\nu}$, as is the case in Ref.\cite{MillerSpheric}.
For information on the Laguerre polynomials see, e.g., Ref.\cite{Gradshteyn}. The first few vector polynomials
are listed in Ref.\cite{MillerSpheric} and they can be obtained up to arbitrary orders by algebraic means. 

The expansion of $\feq$ can be performed as follows:
\begin{equation}
 \feq(t,\vec{x},p,\vec{v}) = \frac{e^{-\bar{p}}}{4\pi T_0^2}\sum_{\ell=0}^{N_p}\sum_{n=0}^{N_v}
 \frac{1}{\ell+1}\,a^{(n\ell)}_{{\rm eq}}(t,\vec{x})\, P^{(n)}_{i_1\dots i_n}(\vec{v}) P^{(n)}_{i_1\dots i_n}(\vec{u})\, L_\ell^{(1)}(\bar{p}),
 \label{feqtrunc}
\end{equation}
where the exact expression of the expansion coefficients $a^{(n\ell)}_{\rm eq}$ is omitted here for brevity.
Truncating the above expansion at order $\ell = N_p$ with respect to the radial component $p$ and at order $n = N_v$
with respect to the angular components $\vec{v}$ ensures the exact recovery of moments of the form:
\begin{equation}
 \int \frac{d^3p}{p^0} \, \feq\, P(p, \vec{v}),\label{mom_feq}
\end{equation}
where $P$ is a polynomial of order at most $N_p$ in $p$ and $N_v$ in $\vec{v}$. The procedure for the recovery of the above moments
is discussed in the following subsection.

\subsection{Discretization of the momentum space using quadrature rules}

To recover the moments given in Eq.~(\ref{mom_feq}), the integrals can be performed using quadrature methods, as follows:
\begin{eqnarray}
&\textrm{\bf a)}&
\textrm{The azimuthal integral:} 
\ 
\ \mathcal{Q}(p,\theta) = \int\limits_0^{2\pi} d\phi\ \feq(p,\theta,\phi) P(p, \vec{v}) =
  \frac{2\pi}{Q_\phi}\sum_{k\,=\,1}^{Q_\phi} \feq(p,\theta,\phi_k) P(p, \theta, \phi_k).
\label{azimuthInt}\\
&\textrm{\bf b)}& \textrm{The polar integral:\phantom{abcd,oo}}
\ 
\mathcal{E}(p) =  \int\limits_{-1}^1 d\xi \ \mathcal{Q}(p,\xi) = \sum_{j\,=\,1}^{Q_\xi} w_{j}^\xi\ \mathcal{Q}(p,\xi_j), \qquad 
w_{j}^\xi = \frac{2(1-\xi_j)^2}{(Q_\xi + 1)^2 \left[P_{Q_\xi+1}(\xi_j)\right]^2}.  \label{polarInt}\\
&\textrm{\bf c)}& \textrm{The radial integral:}
\qquad \qquad \hspace{4pt}
\int d\wp\, \wp\, e^{-\wp}\  \frac{\mathcal{E}(p)}{e^{-\wp}} = \sum_{i\,=\,1}^{Q_p} w_{i}^p\ \frac{\mathcal{E}(p_i)}{e^{-\wp_i}}, \qquad
\quad \hspace{-3pt}
w^p_{i} = \frac{\wp_i}{\left(Q_p + 1\right)\left[L^{(1)}_{Q_p+1}(\wp_i) \right]^2}.
\end{eqnarray}
In the above, $\xi = \cos(\theta)$. The procedure is explained in detail in Refs.\cite{MillerSpheric,VictorSpheric}.\par
The quadrature relations are exact if the integrand in (\ref{azimuthInt}) contains combinations of $\sin\phi$ and 
$\cos\phi$ at combined powers of less
than $Q_\phi$, the function $\mathcal{Q}$ contains powers of $\xi$ of less than $2Q_\xi$ and $\mathcal{E}$ is a 
polynomial in $p$ of order less than $2Q_p$.
For the truncation (\ref{feqtrunc}),
we have the absolute conditions $N_p < Q_p$ and $2N_v < {\rm min}(2Q_\xi,Q_\phi)$ \cite{MillerSpheric,VictorSpheric}. 
The above quadrature rules are valid
for any set of orthogonal polynomials that are defined on
the appropriate domain. We have chosen a simple
trigonometric quadrature for the azimuthal integral \cite{Zwillinger,Mysovskikh88},
Legendre polynomials for the polar angle (Gauss-Legendre quadrature \cite{Abramowitz,hildebrand87}) and Laguerre polynomials for the
radial integral (Gauss-Laguerre quadrature \cite{Abramowitz,hildebrand87}). 
The functions $w^{\,i}_p, w^{\,j}_\xi$ and $w^{\,k}_\phi = 2\pi / Q_\phi$
are called \emph{quadrature weigths}. The discrete set of momenta are 
given by $\phi_k = \frac{k\pi}{Q_\phi}$ while $\xi_j$ and $p_i$
are zeroes of Legendre and Laguerre functions, i.e. 
$L^{(1)}_{Q_p}(p_i) = 0$ and $P_{Q_\xi}(\xi_j) = 0$.
Thus, the complete set of discrete momenta comprises the following elements (in spherical coordinates):
\begin{equation}
\vec{p}_{ijk} = (p_i, \xi_j, \phi_k)\ \ \rightarrow \ \ \vec{p}_s, \ \ \ s = 1\ ...\ n_{\rm vel}.
\end{equation}
The total number of velocities is equal to $n_{\rm vel} = Q_p \times Q_\xi \times Q_\phi$.

\section{RIEMANN PROBLEM}

To test our models, we perform simulations of the Riemann problem with the following initial conditions:
\begin{equation} \label{condinit}
f(z, t = 0, \vec{p}) =
\left\{
\begin{array}{lr}
 {\displaystyle \feq(P_L, n_L, u_L)} & z < 0,\\
 {\displaystyle \feq(P_R, n_R, u_R)} & z > 0,
\end{array}\right.
\end{equation}
where $(P_L,n_L,u_L) = (1,1,0)$ and $(P_R,n_R,u_R) = (0.1,0.125,0)$.
Open boundary conditions are imposed along the axis parallel to the flow (i.e. the $z$ axis) 
and periodic conditions in the perpendicular directions
\cite{gan11}.

In order to determine the quadrature order to be employed for the recovery of the dynamics of $N^\mu$ and $T^{\mu\nu}$, we
consider the projection of Eq.~\eqref{Boltz} on the polynomial $L_\ell^1(\wp)$:
\begin{equation}
 \left(\partial_t + v^z  \partial_z\right) a_\ell = -\frac{u^0 - v^z u^z }{\tau} \left(a_\ell - a_\ell^{\rm eq}\right),
 \label{eq:Boltz_l}
\end{equation}
where $a_\ell$ are the expansion coefficients of $f$ with respect to the Laguerre polynomials, defined through:
\begin{equation}
 f = \frac{e^{-\wp}}{T_0^2} \sum_{\ell =0}^\infty \frac{1}{\ell + 1} a_{\ell} L_\ell^{(1)}(\wp), \qquad
 a_\ell = \int_0^\infty dp\, p\, e^{-\wp} f\, L_\ell^{(1)}(\wp),
\end{equation}
while $a_\ell^{\rm eq}$ are defined in a similar way in terms of $\feq$.
Eq.~\eqref{eq:Boltz_l} shows that the evolution of $a_\ell$ is fully determined by $a_\ell$ and $a_\ell^{\rm eq}$.
Since $N^\mu$ and $T^{\mu\nu}$ are moments of order $1$ and $2$ with respect to $\wp$, respectively, their evolution
is fully determined by the evolution of $a_0$, $a_1$ and $a_2$. Thus, the evolution of $N^\mu$ and $T^{\mu\nu}$ can
be exactly recovered using a quadrature of order
$Q_p = 3$ on the $\wp$ coordinate. Further, the $\phi$ coordinate can only appear through
combinations of $p^x$ and $p^y$. Since Eq.~\eqref{eq:Boltz_l}, as well as the initial conditions \eqref{condinit},
do not contain $p^x$ and $p^y$, the quadrature order along the $\phi$ direction can be taken to be $Q_\phi = 3$.
Furthermore, since we only require the evolution of the $t$ and $z$ components of $T^{\mu\nu}$, 
a quadrature order $Q_\phi = 2$ is sufficient.
The quadrature order $Q_\xi$ along the $v^z = \cos\theta = \xi$ direction is left as a variable which we will use to
control the accuracy of our LB simulations, as will be described in the next section.

\section{NUMERICAL RESULTS}

The discretization of the time-derivative in (\ref{Boltz}) is done using an explicit, nonlinearly stable 3rd order Runge-Kutta
algorithm \cite{Rezzolla}. For the spatial discretization we use a 5th order WENO scheme \cite{Wang07}. It was shown that the
WENO scheme is suitable when simulating flows with discontinuities or strong gradients, in effect suppressing spurious oscillations
and reducing numerical viscosity \cite{gan11}. The spatial domain is set to unity and we chose a grid consisting of
$1 \times 1 \times L_z$ nodes, while the time step is limited by the CFL condition \cite{gan11}.

\begin{figure} \label{Evolution}
\begin{tabular}{ccc}
 \includegraphics[width=0.45\linewidth,height=0.45\linewidth]{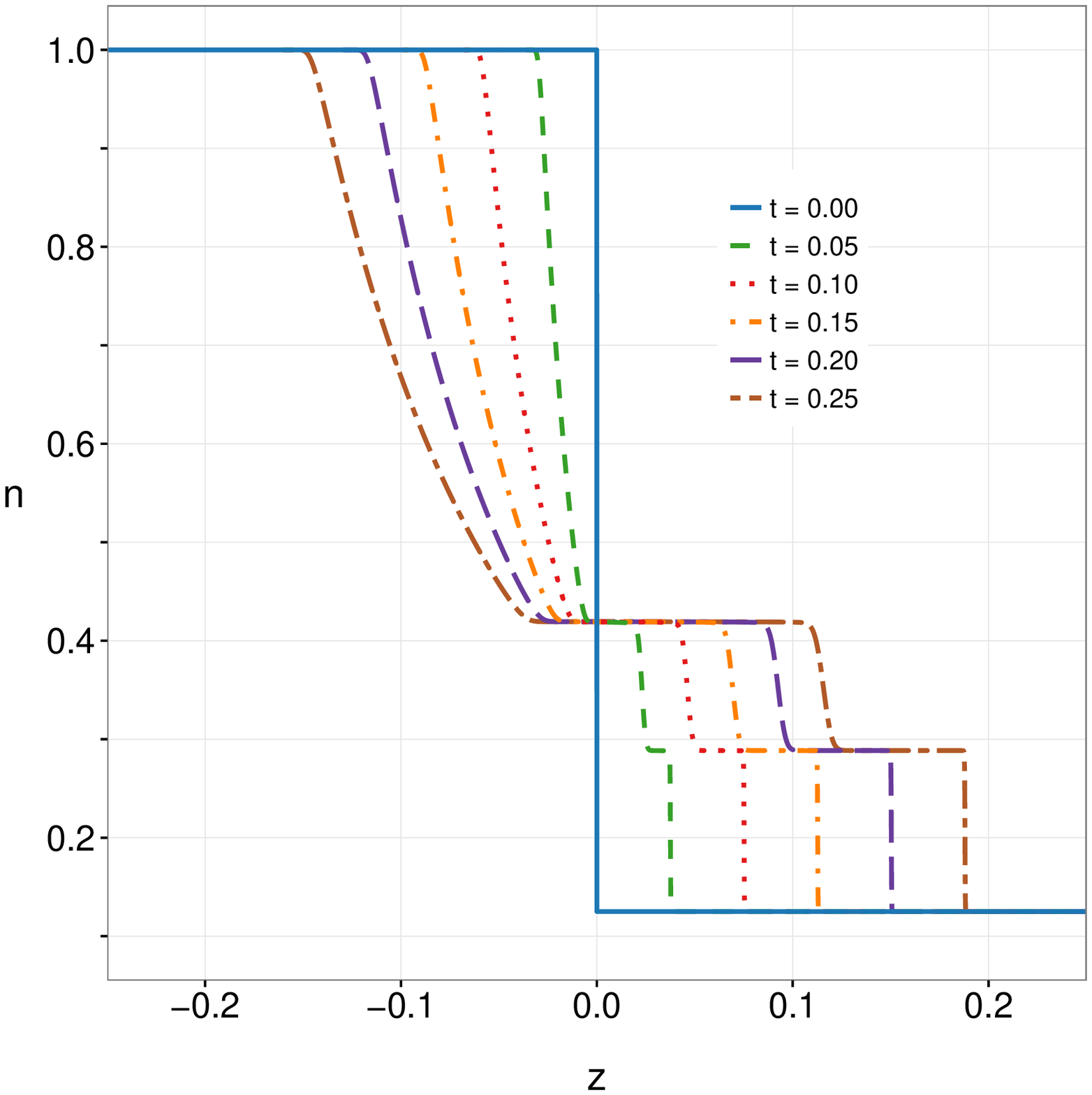} &
 \quad
\qquad &
 \includegraphics[width=0.45\linewidth,height=0.45\linewidth]{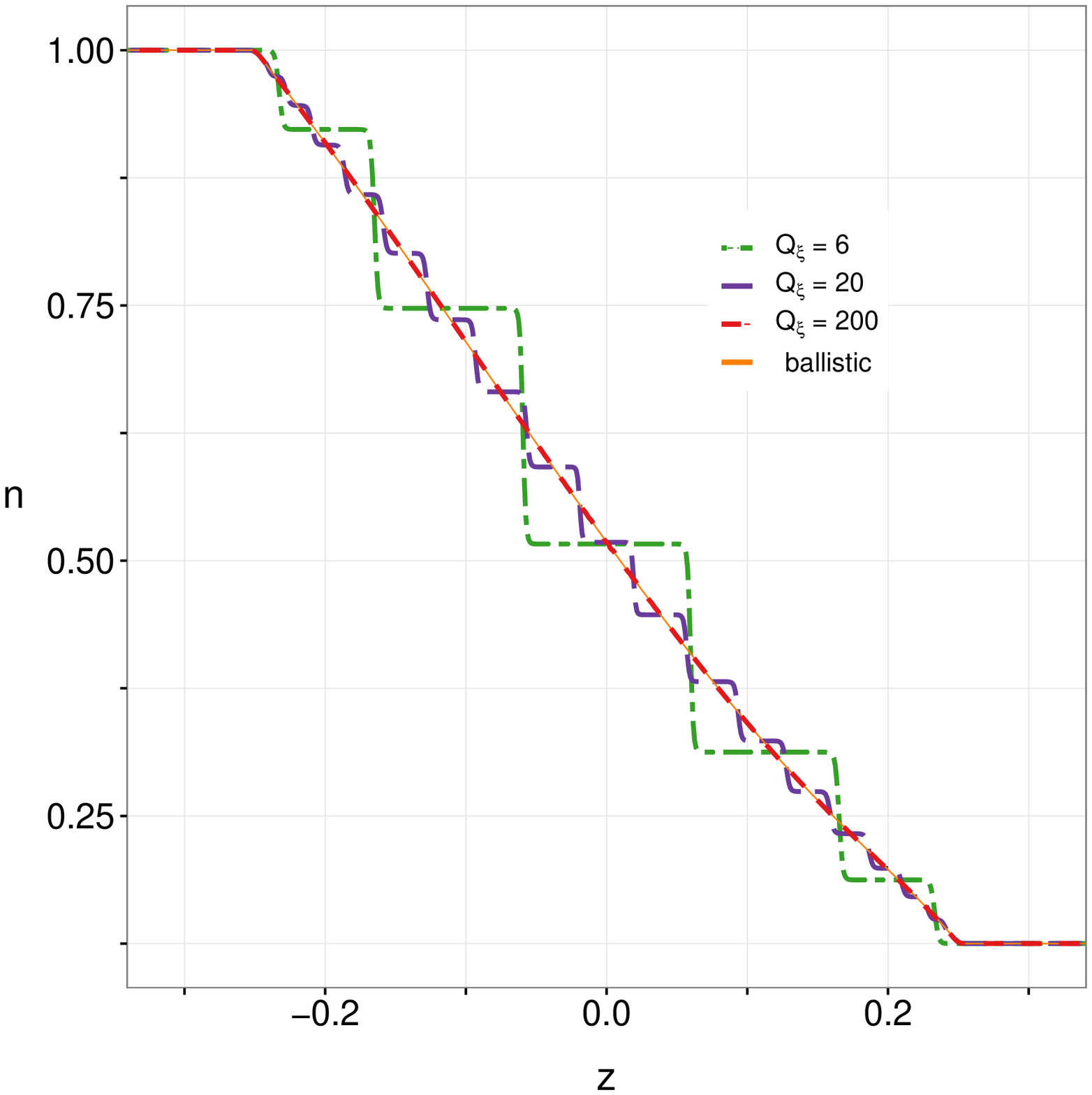}\\
 (a) & & (b)
\end{tabular}
\caption{Density profile for the Riemann problem with initial conditions (\ref{condinit}).
The left panel (a) shows the time evolution of the density in the case of $\tau = 10^{-4}$. In the right panel (b),
we have superimposed the density profiles obtained in the ballistic regime ($\tau \rightarrow \infty$) at a
fixed time
 ($t = 0.25$)
for different values of the quadrature order $Q_\xi$. The profile corresponding to $Q_\xi = 200$ and 
the analytic solution \eqref{ballistic} are overlapped.}
\end{figure}

In the {\bf hydrodynamic regime}, the Riemann problem is well studied \cite{Marti94,Rezzolla06}. The results of our
simulations for the initial conditions \eqref{condinit} with $\tau = 10^{-4}$, time-step $\delta t = 0.5 \times 10^{-4}$
and a number of $L_z = 10000$ nodes along the $z$ axis can be seen in Fig.~\ref{Evolution}(a).
As the fluid evolves, we observe the formation of a rarefaction wave, a contact discontinuity and a shock-wave. In ideal hydrodynamics,
the shock-wave represents a discontinuity in the density and pressure. As is the case in all BGK-type LB models, the viscosity depends
linearly on the relaxation time $\tau$.
Thus, in numerical simulations with a finite $\tau$, the shock (and contact discontinuity) will always
become smoothed as a consequence of this non-vanishing viscosity. The width of the shock-wave is a good measure of the numerical accuracy of
the model. In our case, the width is equal to around 6-7 grid nodes, which represents a distance of $\Delta z \sim  6 \times 10^{-4}$.
To obtain the aforementioned results, we used $N_p = 2$ and $N_v = 5$ in Eq.~\eqref{feqtrunc} and $Q_\xi = 6$.

\begin{figure}[t] \label{Rconv}
\includegraphics[width=0.45\linewidth,height=0.45\linewidth]{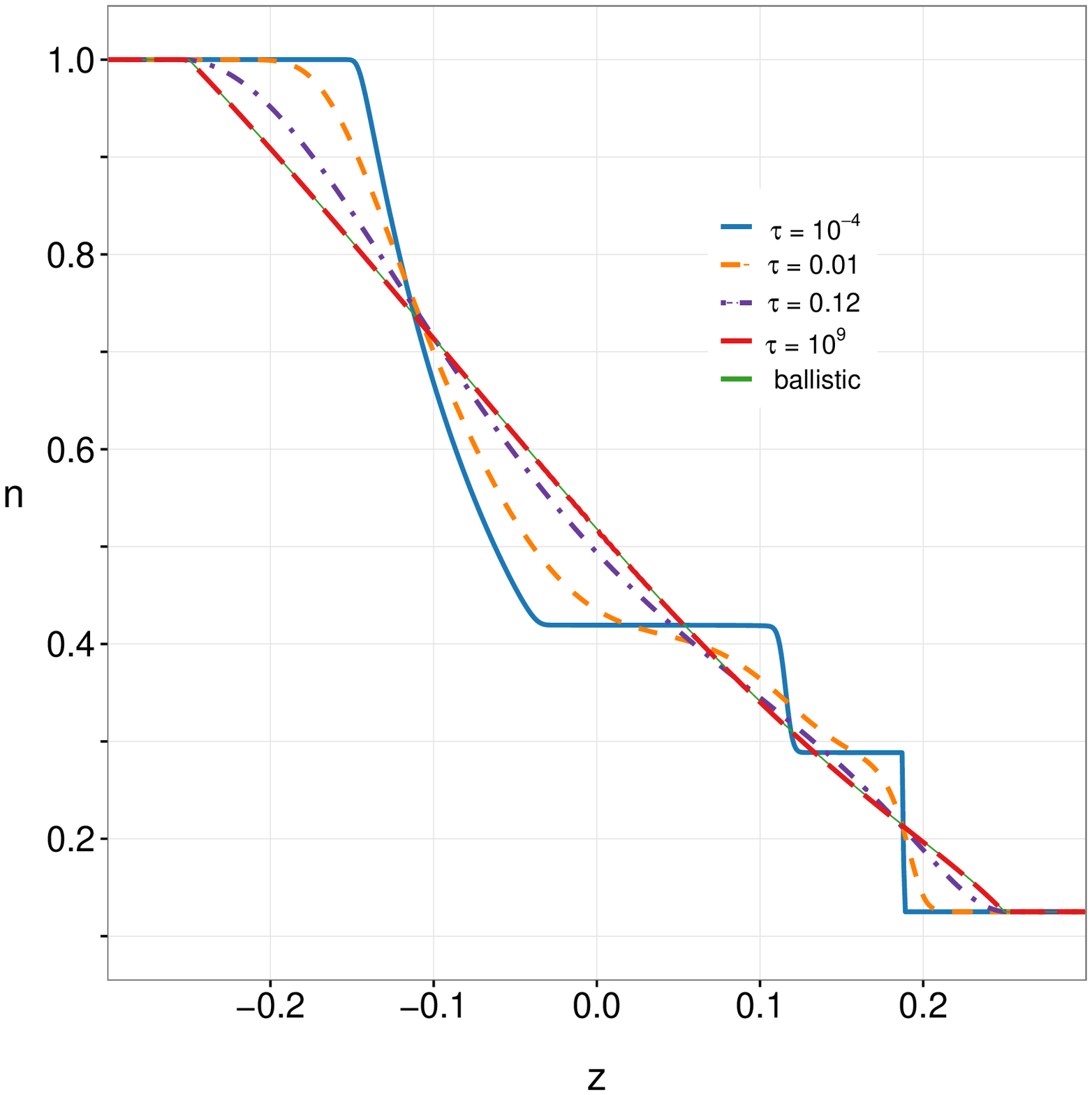}
\quad
\qquad
\includegraphics[width=0.45\linewidth,height=0.45\linewidth]{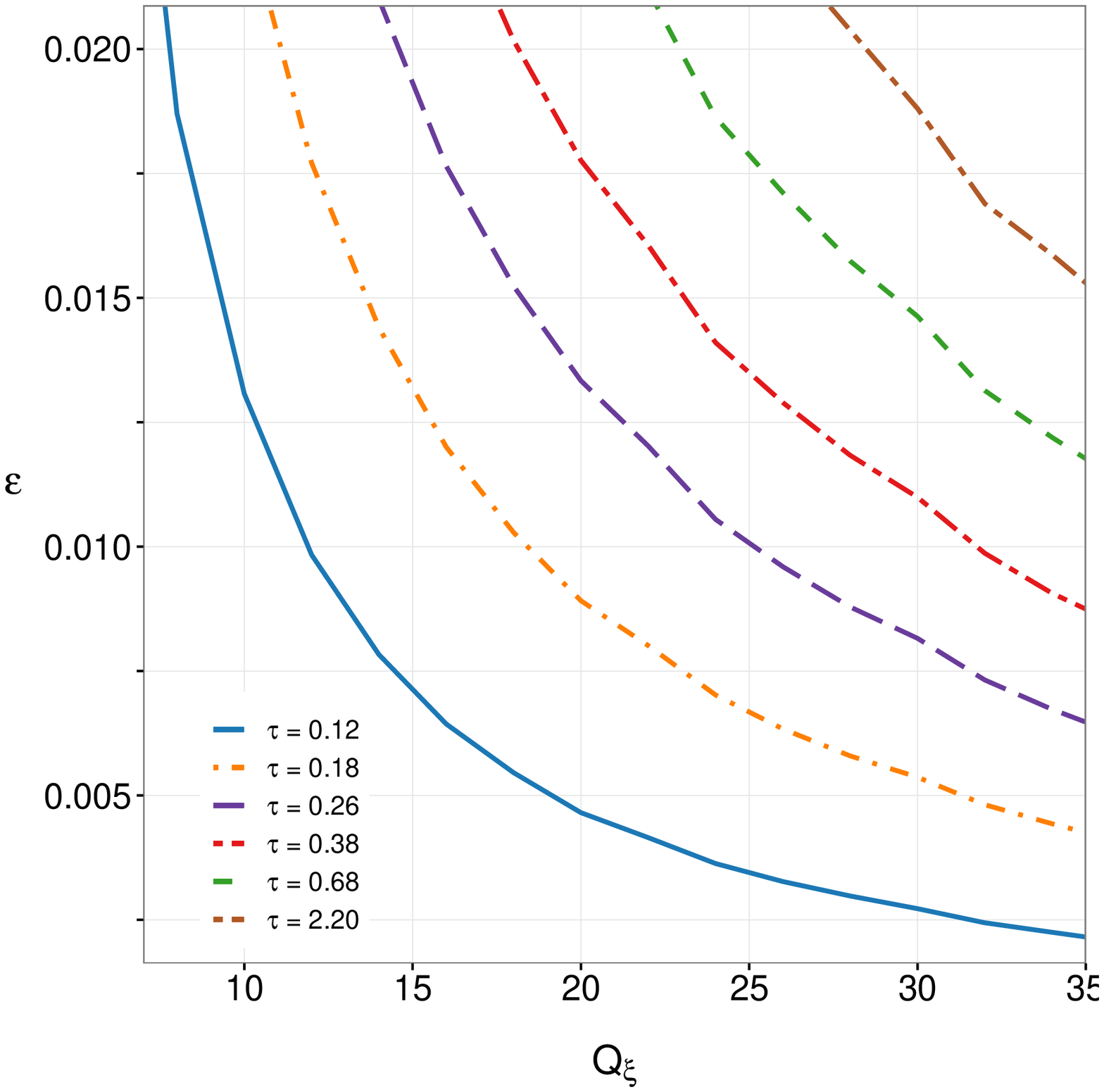}
\caption{(a) Density profile at
 $t = 0.25$
for various values of $\tau$.
(b) Dependence of the error \eqref{eq:err} with respect to the quadrature procedure for
various values of $\tau$. The reference profiles were obtained using $Q_\xi = 200$.}
\end{figure}

In the {\bf ballistic regime} ($\tau \rightarrow \infty$), the right hand-side of the Boltzmann equation vanishes, and an
analitic solution of the equation can be found. For the ansatz (\ref{condinit}), the density profile in the ballistic regime
is given by:
\begin{equation} \label{ballistic}
n_{\rm bal} = \left\{
                \begin{array}{lcr}
                  n_L & \qquad & z < -t, \\
                  {\displaystyle \sqrt{\left(\frac{n_L + n_R}{2} - \frac{n_L-n_R}{2}\,\frac{z}{t} \right)^2 - \,
                  \left(\frac{n_L - n_R}{4}\right)^2\left(1-\frac{z^2}{t^2}\right)^2}} & \qquad & -t < z < t, \\
                  n_R & \qquad & z > \,t.
                \end{array}
              \right.
\end{equation}
Figure~\ref{Rconv}(b) shows the convergence trend of our simulation results as $Q_\xi$ is increased.
The agreement between the case when $Q_\xi = 200$ and the above analytic result is excellent.

With our model, we can also capture the evolution of the fluid at {\bf finite relaxation times}, where viscous and rarefaction effects
become significant. In Fig.~\ref{Rconv}(a), we have represented the density profile of the fluid at time
 $t = 0.25$, for different
values of the relaxation time. As $\tau$ increases, and thus the system becomes more dissipative, the features are smoothed.
As we go towards increasing values of the relaxation time $\tau$, increasing quadrature orders are required for the
recovery of the physics of the flow. In order to asses the capability of a model of given $Q_\xi$ to simulate the
Riemann problem, we considered the following quantity \cite{ambrus16a,ambrus16b}:
\begin{equation}
 \varepsilon = {\rm max}{z} \left[\frac{n(z) - n_{\rm ref}(z)}{\Delta n}\right],
 \label{eq:err}
\end{equation}
where $\Delta n = n_L - n_R = 0.875$ and $n_{\rm ref}(z)$ is a ``reference'' density profile. In the absence of an
analytic expression for $n_{\rm ref}$, we have considered the profile obtained using a large quadrature order,
i.e.~$Q_\xi = 200$.
The quantity $\varepsilon$ thus represents the maximum relative deviation of the profile $n(z)$ obtained with a
model of a given $Q_\xi$ from the reference profile $n_{\rm ref}(z)$, obtained using $Q_\xi = 200$.
Figure~\ref{Rconv}(b) shows the dependence of $\varepsilon$ on $Q_{\xi}$ for various values of $\tau$. It can be
seen that the quadrature order required to reduce $\varepsilon$ under the $1\%$ threshold (where we consider that
{\em convergence} is achieved) increases as $\tau$ increases. In the hydrodynamic regime ($\tau = 10^{-4}$),
we found that $Q_\xi = 4$ is sufficient to achieve convergence.
All results presented in Fig.~\ref{Rconv} were obtained using $N_p = 2$ and $N_v = 5$.

\section{CONCLUSIONS}
We have developed a quadrature-based lattice Boltzmann model for simulating relativistic flows. The model was tested
on a version of the classical Riemann problem (i.e.~the Sod shock tube). The results obtained in the hydrodynamic regime
are consistent with those in the literature, while in the ballistic regime, we show that our models recover the analytic
result. In order to asses the accuracy of our models for finite values of the relaxation time $\tau$, we have considered
a convergence test which requires the relative error with respect to some reference profile obtained using
a $200$-point quadrature to be below $1\%$. We have shown that,
in the hydrodynamic regime, a small order of the quadrature is sufficient to obtain convergence, while in the ballistic regime,
the convergence is slow. The ease with which our model can be extend to arbitrary orders makes it a pragmatic tool for simulating
flows across the whole spectrum of relaxation times.

{\bf Acknowledgement.} This work was supported by a grant of the Romanian National Authority for Scientific Research and Innovation, CNCS-UEFISCDI, project number
PN-II-RU-TE-2014-4-2910.

\end{document}